\documentclass[]{aastex63}
\usepackage{amsmath,amsfonts,amssymb}
\usepackage{graphicx}
\hypersetup{colorlinks=true, allcolors=blue} 
\shorttitle{2$\%$ Distance to M31}
\shortauthors{Li et al.}
\graphicspath{{./}{figures/}}

\begin{document}

\title{A sub-2$\%$ Distance to M31 from Photometrically Homogeneous Near-Infrared\\ Cepheid Period-Luminosity Relations Measured with the \textit{Hubble Space Telescope}}

\correspondingauthor{Siyang Li}
\email{sli185@jhu.edu}

\author[0000-0002-8623-1082]{Siyang Li}
\affiliation{Department of Physics and Astronomy, Johns Hopkins University, Baltimore, MD 21218, USA}

\author[0000-0002-6124-1196]{Adam G. Riess}
\affiliation{Department of Physics and Astronomy, Johns Hopkins University, Baltimore, MD 21218, USA}
\affiliation{Space Telescope Science Institute, 3700 San Martin Drive, Baltimore, MD 21218, USA}

\author[0000-0003-4961-6511]{Michael P. Busch}
\affiliation{Department of Physics and Astronomy, Johns Hopkins University, Baltimore, MD 21218, USA}

\author{Stefano Casertano}
\affiliation{Space Telescope Science Institute, 3700 San Martin Drive, Baltimore, MD 21218, USA}

\author[0000-0002-1775-4859]{Lucas M. Macri}
\affiliation{Department of Physics and Astronomy, Texas A$\&$M University, College Station, TX 77845, USA}

\author[0000-0001-9420-6525]{Wenlong Yuan}
\affiliation{Department of Physics and Astronomy, Johns Hopkins University, Baltimore, MD 21218, USA}

\nocollaboration{6}



\begin{abstract}

We present Period-Luminosity Relations (PLRs) for 55 Cepheids in M31 with periods ranging from 4 to 78 days observed with the {\it Hubble Space Telescope} (HST) using the same three-band photometric system recently used to calibrate their luminosities. Images were taken with the Wide Field Camera 3 in two optical filters (F555W and F814W) and one near-infrared filter (F160W) using the Drift And SHift (DASH) mode of operation to significantly reduce overheads and observe widely-separated Cepheids in a single orbit. We include additional F160W epochs for each Cepheid from the Panchromatic Hubble Andromeda Treasury (PHAT) and use light curves from the Panoramic Survey Telescope and Rapid Response System of the Andromeda galaxy (PAndromeda) project to determine mean magnitudes. Combined with a 1.28$\%$ absolute calibration of Cepheid PLRs in the Large Magellanic Cloud from \cite{Riess_2019} in the same three filters, we find a distance modulus to M31 of $\mu_0$ = 24.407 $\pm$ 0.032, corresponding to 761 $\pm$ 11 kpc and 1.49$\%$ uncertainty including all error sources, the most precise determination of its distance to date. We compare our results to past measurements using Cepheids and the Tip of the Red Giant Branch (TRGB).  This study also provides the groundwork for turning M31 into a precision anchor galaxy in the cosmic distance ladder to measure the Hubble constant together with efforts to measure a fully geometric distance to M31. 

\end{abstract}



\section{Introduction} 

Cepheids are variable giant stars that lie in the instability strip of the Hertzsprung--Russell diagram.  They pulsate in a periodic expansion and contraction, oscillations which overshoot their hydrostatic equilibrium points due to the temperature dependence of their atmospheric opacity.  The timescale of these pulsations is proportional to their densities, which are a function of mass and luminosity. The resulting Period--Luminosity relations \citep[PLRs;][]{Leavitt_1912HarCi.173....1L} and their great luminosities allow Cepheids to be used as standard candles to determine extragalactic distances. By first using high-precision geometric distance measurements to calibrate nearby Cepheids in galaxies such as the Milky Way (MW) or the Large Magellanic Cloud (LMC), the relative apparent magnitudes between nearby and more distant Cepheids can be used to determine absolute distances.  In particular, Cepheids play a crucial role in the cosmic distance ladder used to measure the Hubble constant, $H_0$, as Cepheids can be combined with other standard candles such as Type Ia supernovae to determine distances to further galaxies in the Hubble--Lema\^{i}tre flow \citep{Riess_2016}.


The Andromeda galaxy (M31, NGC$\,$224) is the nearest spiral to our own and together with its Cepheid variable V1 played a prominent role in establishing the distance scale of the universe \citep{Hubble_1929}. Due to its proximity, M31 could serve as an anchor in the cosmic distance ladder. However, the lack of a robust geometric calibration, and the relatively high inclination and accompanying extinction, has historically made it difficult to reduce the uncertainty in its distance measurements. Blending has been shown to bias ground-based observations by up to 0.2 mag \citep{Mochejska_2000AJ....120..810M, Vilardell_2007refId0}, and both crowding and differential extinction cause high dispersion in the Cepheid PLR. While past studies \citep{Macri_2001ApJ...549..721M,Riess_2011, Wagner-Kaiser_2015_10.1093/mnras/stv880, Kodric_2018b} have shown that space-borne observations can help reduce the effects of crowding and blending, inhomogeneities between the filter systems used to observe Cepheids in M31 and their geometric calibrations yielded uncertainties in the distance measurements between 3$\%$ and 5$\%$, limiting our knowledge of the distance to M31 and studies which use M31 to calibrate stellar luminosities.

M31 is also an ideal laboratory to investigate differences among different distance measurement methods. Recent measurements of $H_0$ using Cepheid and Tip of the Red Giant Branch (TRGB) methods yield mildly different results that could be due to chance or differences in calibration \citep{Riess_2021, Freedman_2019}. Comparing a high-precision Cepheid measurement to M31 to past TRGB measurements along with other hosts in common may help clarify whether this difference is caused by chance,  differences in calibration or of environments.  

Previously, the large observatory overheads and narrow field of view (FOV) of the \textit{Hubble Space Telescope (HST)} limited the possibility of observing a large number of Cepheids in M31 in the same 3 filters as \cite{Riess_2019}, who recently established a high-precision Cepheid anchor in the LMC based on a robust geometric distance estimate \citep{Pietrzynski_2019}. The availability of the Drift And SHift \citep[DASH;][]{Momcheva_2017} mode of observing starting from Cycle 24 in 2016 allowed us to circumvent these obstacles and observe a greater number of Cepheids (up to 12 Cepheids in 3 filters in one orbit). DASH uses \textit{HST} gyroscopes for guiding and is advantageous for our short integrations of $10-21$s, allowing us to eliminate the 6-minute overhead for each guide star acquisition required for slews beyond $2\arcmin$ under Fine Guidance Sensor (FGS) control. We further used subarray observations to reduce additional overheads that would otherwise require a 350s memory buffer dump for each full UVIS frame. The expected gyro drift was $\sim 0.002\arcsec$ per second, smaller over each exposure than the $0.04\arcsec$ and $0.128\arcsec$ pixels for WFC-UVIS and WFC3-IR, respectively. Further advantages of the DASH mode for observing Cepheids are described in \cite{Riess_2019}.  For each Cepheid we made two observations separated by a week to improve precision and to test the expected phase of the Cepheid.

To augment the number of F160W epochs, we drew upon frames from the Panchromatic Hubble Andromeda Treasury (PHAT) (PI: J. Dalcanton). PHAT is a {\it HST} multi-cycle program that resolved more than 100 million stars in a third of the star-forming disk in M31 using 6 filters ({\it F275W, F336W, F475W, F814W, F110W,} and {\it F160W}) ranging from the ultraviolet to near-infrared \citep{Dalcanton_2012ApJS..200...18D}. We only used the PHAT {\it F160W} measurements since the UV/optical observations were obtained with a different camera (ACS). Due to the relatively small amplitudes of Cepheid light curves in the near-infrared, the {\it F160W} magnitudes at random-phase observations from PHAT are on average close to those at mean phase \citep{Madore_1991PASP..103..933M}. As PHAT uses the same {\it F160W} filter as used in \cite{Riess_2019} and this study, we can use PHAT images to further increase the number of epochs in {\it F160W} and obtain a better estimate of the mean magnitude for each phase-corrected Cepheid. The Panoramic Survey Telescope and Rapid Response System (Pan-STARRS1) survey of the Andromeda galaxy (PAndromeda) obtained light curves for 2686 Cepheids in the {\it gri} bands. Combined with random-phase observations of corresponding {\it HST} Cepheids, the mean magnitude estimations for the Cepheids can be improved using these light curves and previously-constructed templates \citep{Yoachim_2009, Inno_2015_refId0}.

In this study, we use phase-corrected PLRs from multi-epoch {\it HST} observations of 55 M31 Cepheids and a calibration to the LMC that uses the same photometric system and instrument to reduce the uncertainty in the distance to M31 to 1.49$\%$. In \S\ref{sec:CephStandards}, we describe the measurement and photometry of the Cepheids analyzed in this study. In \S\ref{sec:PLR}, we construct PLRs in {\it F555W, F814W, F160W}, and both visible (VIS) and near-infrared (NIR) Wesenheit indices, and in \S\ref{sec:Distance} we use the NIR Wesenheit PLR to derive a distance to M31, compare our result with those in past literature, and discuss the potential use of M31 as a new anchor galaxy. Finally, in \S\ref{sec:Discussion}, we discuss the implications of our results.

\begin{longrotatetable}

\begin{deluxetable}{ccccccccc}
\tablecaption{Observations of M31 Cepheids}

\tablehead{\multicolumn{1}{p{3cm}}{\centering Target \\ (1)} & \multicolumn{1}{p{2cm}}{\centering Frame (idit)\\ (2)} & \multicolumn{1}{p{1cm}}{\centering Filter\\ (3)} & \multicolumn{1}{p{2cm}}{\centering MJD\\ (4)} & \multicolumn{1}{p{2.5cm}}{\centering Exp. Time (s)\\ (5)} & \multicolumn{1}{p{1cm}}{\centering X\\ (6)} & \multicolumn{1}{p{1cm}}{\centering Y\\ (7)} & \multicolumn{1}{p{2cm}}{\centering Program\\ (8)} & \multicolumn{1}{p{1cm}}{\centering Flag\\ (9)}}
\startdata
GRP-11.20822+41.44917 & 510sbq & F555W & 58135.23621 & 10.0 & 897.7 & 471.0 & GO15146 &  \\
\nodata & 510scq & F814W & 58135.23765 & 10.0 & 889.7& 469.7 & GO15146 &  \\
\nodata  & 510spq & F160W & 58135.25567 & 21.3 & 392.3 & 274.9 & GO15146 &  \\
GRP-11.24332+41.50926 & 508g0q & F555W & 58103.65339 & 10.0  & 746.0 & 518.9 & GO15146 &  \\
\nodata  & 507rlq & F814W & 58097.56431 & 10.0  & 711.2 & 523.97 & GO15146 & 
\enddata
\tablecomments{Description of flags for column 9: (1) Excessive drift, PSF semi-major axis greater than 3 pixels for {\it F555W} and {\it F814W} and 2 pixels for {\it F160W}; (2) Cepheid is not fully in frame; (3) Cepheid matching algorithm from \cite{Riess_2019} has fewer than 10 matches and the Cepheid could not be found visually; (4) Cepheid removed because all images in one or more filters were unrecoverable due to flags 1-3.}
\tablecomments{Only a portion of this table is shown here to demonstrate its form and content. A machine-readable version of the full table is available.}
\label{tab:frames}
\end{deluxetable}

\begin{deluxetable}{cccccccccccccccc}
\tablecaption{Photometric Results for M31 Cepheids\label{tab:PhotResults}}
\tabletypesize{\footnotesize}
\tablehead{\colhead{Cepheid} & \colhead{R.A.} & \colhead{Decl.} & \colhead{Period} & \colhead{F160W} & \colhead{$\sigma$} & \colhead{F814W} & \colhead{$\sigma$} & \colhead{F555W} & \colhead{$\sigma$} & \colhead{$m^{W}_{H}$} & \colhead{$\sigma$} & \colhead{$m^{W}_{I}$} & \colhead{$\sigma$} & \colhead{Geo. Corr.} & \colhead{Sample}}
\startdata
CEPH-10.91809+41.18565 & 10.91808 & 41.18564 & 55.938 & 17.002 & 0.057 & 18.757 & 0.057 & 20.841 & 0.085 & 16.213 & 0.098 & 16.048 & 0.170 & 0.019 & Gold\\
CEPH-11.00794+41.50202 & 11.00786 & 41.50202 & 22.200 & 17.997 & 0.062 & 18.987 & 0.057 & 20.298 & 0.069 & 17.507 & 0.099 & 17.283 & 0.159 & 0.002& Gold \\
CEPH-11.02407+41.34854 & 11.02399 & 41.34855 & 17.605 & 18.245 & 0.072 & 19.677 & 0.100 & 21.343 & 0.146 & 17.617 & 0.121 & 17.512 & 0.298 & 0.014 & Gold\\
CEPH-11.05340+41.30685 & 11.05333 & 41.30685 & 10.631 & 19.141 & 0.088 & 20.493 & 0.223 & 22.175 & 0.339 & 18.507 & 0.192 & 18.307 & 0.676 & 0.018 & Gold\\
CEPH-11.09119+41.37223 & 11.09104 & 41.37233 & 13.156 & 18.615 & 0.079 & 19.713 & 0.168 & 21.046 & 0.191 & 18.116 & 0.144 & 17.981 & 0.460 & 0.016 & Gold
\enddata
\tablecomments{Only a portion of this table is shown here to demonstrate its form and content. A machine-readable version of the full table is available. Gold sample errors are calculated by adding the photometric and template errors in quadrature. An intrinsic scatter of 0.069 mag from \cite{Riess_2019} is added to the NIR Wesenheit in quadrature after propagation. Silver sample F555W, F814W, and F160W mean magnitude errors are calculated by adding in quadrature the photometric errors for each epoch and dividing by the square root of the number of epochs used. These errors are propagated into the VIS and NIR Wesenheit indices. We add a random phase error of 0.13 mag from \cite{Macri_2015AJ....149..117M} in the VIS and NIR Wesenheits in quadrature with the propagated uncertainties as well as the intrinsic scatter of 0.069 mag to the NIR Wesenheit index. Geometric corrections have been added to the VIS and NIR Wesenheits magnitudes in this table.}
\end{deluxetable}
\end{longrotatetable}

\section{M31 Cepheid Standards}
\label{sec:CephStandards}

\subsection{DASHing through M31} \label{sec:style}

We obtained multi-epoch \textit{HST} images of 57 Cepheids in M31 in two WFC3-UVIS optical filters ({\it F555W, F814W}) and one WFC3-IR near-infrared filter ({\it F160W}) as part of program GO-15146 (PI: Riess) using DASH. Exposure times were 10s for {\it F555W} and {\it F814W} and 21.3s for {\it F160W}. We intended to obtain two epochs per filter for each Cepheid, but \textit{HST} Gyro 2 experienced severe degradation during our observations, resulting in sporadic and unpredictable drift --- in some cases exceeding its expected value by more than an order of magnitude. Characterization of this degradation can be found in \cite{Riess_2019}. As a result, 45 out of the 392 images obtained were unusable, leaving 55 Cepheids for analysis.

\begin{figure}[ht!]
\plotone{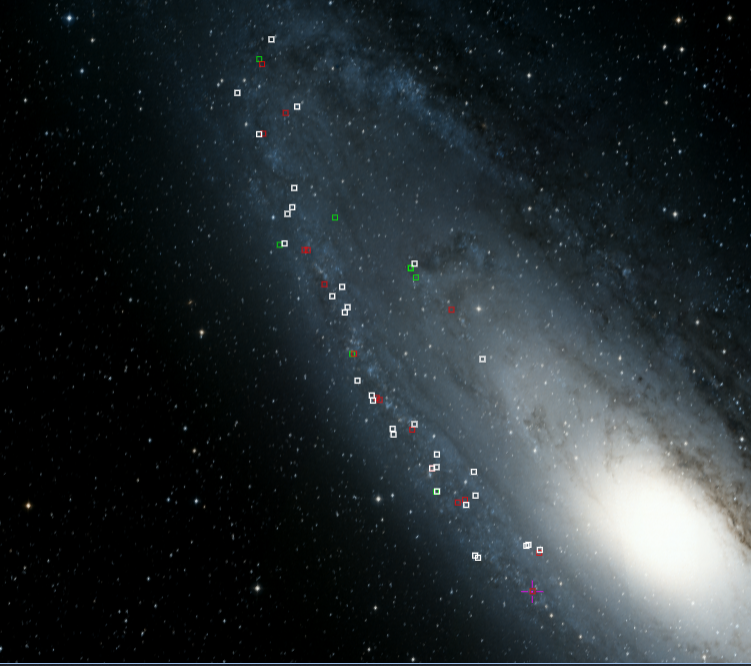}
\caption 
{Locations of the 55 Cepheids analyzed in this study. Mean magnitudes residuals of $e>= 0.5, -0.5 < e < 0.5,$ and $e<= 0.5$ using the {\it F555W} PLR fit from \cite{Riess_2019} are shown in red, white, and green, respectively. Image generated using AstroView in the Mikulski Archive for Space Telescopes website.}
\label{fig:CephLocations}
\end{figure}

Fig. \ref{fig:CephLocations} shows the locations of these Cepheids within M31, while Table \ref{tab:frames} lists the following properties: (1) target ID, (2) corresponding frame, (3) filter, (4) date of observation, (5) exposure time, (6) image X coordinates, (7) image Y coordinates, (8) corresponding program (GO-15146 or PHAT), and (9) image removal flag (described in the table footnote).

We removed CEPH-11.01715+41.46200 and CEPH-11.21898+41.47505 from further consideration due to flag 4 (see Table~\ref{tab:frames}), and further determined that two Cepheids in GRP-11.09581+41.35417 (P  = 36.106, 36.154) and two Cepheids in GRP-11.29561+41.60411 (P = 48.447, 48.472 days) were the same object. For each of these two Cepheids, we combine their measurements and take the average of the two periods for each pair in our analysis. The final sample consists of 55 unique variables. 

\subsection{Photometry}
\label{sec:Photometry}

We retrieved calibrated frames using the Mikulski Archive for Space Telescopes (MAST) website: {\tt flc} suffix for the optical images, corrected for charge transfer efficiency (CTE), and {\tt flt} suffix for the near-infrared ones. The goal of GO-15146 was to obtain two epochs in each filter for each Cepheid separated by one week to increase the accuracy of the mean magnitude recovered using template fitting (described in \S\ref{sec:PhaseCorrSection} below). We used ground-based $r$-band light curves from PAndromeda \citep{Kodric_2018} for phase corrections, since that band had the greatest number of epochs and frames per epoch on average available for each Cepheid. We find an average of 319 epochs per Cepheid for the PAndromeda lightcurves used in this study. We created three samples of Cepheids: ``gold", ``silver", and ``gold+silver". The gold sample consists of 42 Cepheids having $r$-band light curves from PAndromeda that can be used for phase corrections. The silver sample consists of 13 Cepheids that either do not have corresponding light curves or had light curves rejected by \cite{Kodric_2018} using a color cut. In the silver sample we adopt an unweighted average of the epoch magnitudes of each Cepheid to avoid bias that would be caused by higher statistical uncertainties in fainter Cepheids as well as the larger uncertainty due to their random phase.  The gold+silver sample combines the gold and silver samples. 

Because the use of DASH resulted in an inaccurate World Coordinate System for each GO-15146 image, we first located Cepheids using the algorithm from \cite{Riess_2019} and centered the Cepheid using DAOStarFinder from the {\tt photoutils} package in Python \citep{bradley_2020_4049061}. We measured all magnitudes using aperture photometry with PythonPhot, which uses the DAOPHOT routines \citep{DAOPHOT_1987PASP...99..191S} translated into Python by \cite{Jones_2015}. Aperture photometry, with radii much larger than the point spread function (PSF) FWHM and the measured field drift, is advantageous over PSF fitting for the frames used in this study because it is less sensitive to the broadening of the PSF that could result from gyroscopic drift. This drift was measured from the change in the positions of the intended targets between successive images.
 
We multiplied each frame by a pixel area map, provided by STScI, corresponding to the respective camera and position of the subarray. The pixel area maps are used to correct for the photometric impact of the flat-fielding process in the CALWF3 pipeline on pixels of non-uniform size. We set the aperture radii to 3 pixels ($\sim 0.12\arcsec$) for WFC3-UVIS images and 2 pixels ($\sim 0.25\arcsec$) for WFC3-IR images. While smaller aperture radii can increase the nominal signal-to-noise ratio, they also increase sensitivity to PSF broadening. We selected these aperture radii to balance these two effects, as well as to reduce contamination from nearby stars or cosmic rays. We tested for potential errors from large drifts, flagging and removing cases where the error was not negligible.

For images taken in {\it F555W}, we applied a zeropoint of 25.727 mag (Vega system) and an aperture correction of $-0.183$ mag from a 3-pixel radius to infinity. For images taken in {\it F814W}, we applied a zeropoint of 24.581 mag and an aperture correction of $-0.272$ mag from a 3-pixel radius to infinity. These values are consistent with those used by \cite{Riess_2019} for LMC Cepheids. For images taken in {\it F160W}, we applied a zeropoint of 24.71 mag and the same aperture correction of $-0.200$ mag from a 3-pixel radius to infinity derived by \cite{Riess_2019}, combined with a $-0.166$ mag correction from a 2-pixel to a 3-pixel radius for a total aperture correction of $-0.366$ mag from a 2-pixel radius to infinity. Although this zeropoint may differ by $\sim$ 0.02 mag from the latest or future STScI calibrations or from small WFC3-IR sensitivity variations over time \citep{Kozhurina_2020wfc..rept....5K}, we used the same values used in \cite{Riess_2019} to nullify zeropoint errors through the comparison to LMC Cepheids.

In a few cases, visual inspection indicated the presence of significant crowding near the target Cepheid. For these objects, we used smaller aperture radii of 2 pixels for {\it F555W} and {\it F814W} images and 1 pixel for {\it F160W} images, along with corresponding aperture corrections to infinity of $-0.450, -0.500$ and $-0.944$~mag. Although applying such smaller apertures decreases the signal-to-noise ratio, it further reduces contamination from nearby stars that would bias our photometry for these particular targets.

\subsection{Phase Corrections}
\label{sec:PhaseCorrSection}

\begin{figure}[ht!]
\plotone{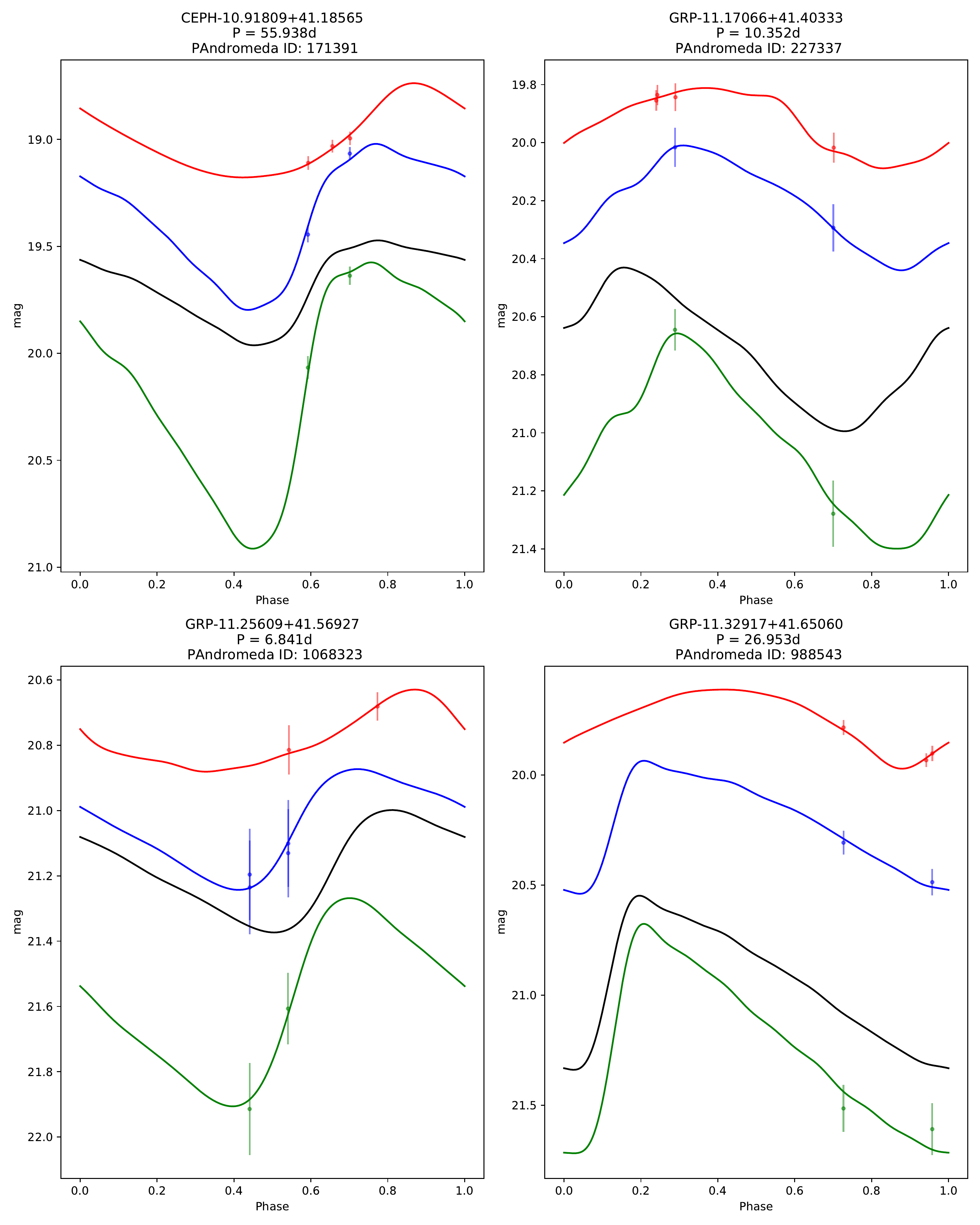}
\caption 
{Four example template fits from the gold sample. Black lines correspond to $r$-band lightcurves from PAndromeda. Red, blue, and green correspond to {\it F160W, F814W, and F555W}, respectively. Dots with error bars represent \textit{HST} observations and their photometric errors while solid lines represent the template fit. Curves are shifted as follows: upper left: {\it F160W}+2, {\it F814W}+0.6, {\it F555W}-0.7, upper right:  {\it F160W}+1, {\it F814W}+0.2, lower left:  {\it F160W}+1.2, {\it F814W}+0.4, {\it F555W}-0.1, lower right:  {\it F160W}+2, {\it F814W}+0.8,{\it F555W}-0.2.}
\label{fig:PhaseCorr}
\end{figure}

We estimate the mean magnitude for each Cepheid in the gold sample from our random multi-epoch observations by applying phase corrections using periods and ground-based light curves from PAndromeda \citep{Kodric_2018}. Using PAndromeda light curves to estimate mean magnitudes for the gold-sample Cepheids results in lower uncertainties compared to silver-sample Cepheids, which use an unweighted mean for the magnitude estimate. The additional information provided by the PAndromeda light curves allows us to infer the amplitude and phase of the gold-sample Cepheids, which is not possible for the silver sample. Figure \ref{fig:PhaseCorr} shows examples of template fits for four Cepheids. We use $V$- and $I$-band templates from \cite{Yoachim_2009} and $H$-band templates from \cite{Inno_2015_refId0} and approximate the $V$, $I$, and $H$ bands as {\it F555W}, {\it F814W}, and {\it F160W}, respectively. The uncertainties in the color terms used to transform between the {\it HST} and Johnson-Cousins filter systems are negligible. \cite{Riess_2019} find that a 10$\%$ error in the color term would change the transformed magnitudes by less than 0.1 mmag. The procedure we use simultaneously fits templates to measurements in all three filters and has five free parameters: phase, $V$-band amplitude, and three mean magnitudes. We fix the ratio between the $H$ and $V$ band amplitudes to 0.34 for $P\le 20$~d and 0.40 for $P>20$~d following \cite{Inno_2015_refId0} and the ratio between $I$ and $V$ band amplitudes to 0.58 following \cite{Yoachim_2009}. Using a single $H$-to-$V$ amplitude ratio of 0.40 mag does not change the final distance. The template-fitting procedure performs a two-parameter grid search through phase and $V$-band amplitude and estimates a mean magnitude correction for each filter using the Modified Julian Date, period, phase, and amplitude at each iteration. This correction is added to the {\it HST} epochs and is not a function of magnitude and so is constant for a given combination of period, filter, phase, and amplitude. This correction is then added to each epoch to estimate the mean magnitude, and if there is more than one epoch for a given filter, we take a weighted average of the estimated mean magnitudes using the photometric errors as weights. For each iteration through the grid search, the chi squared, $\chi^2$, is computed for all three filters and $V$-band amplitude with the estimated mean magnitudes used as the expected value and the {\it HST} epochs used as the actual value in the $\chi^2$ terms. As the $V$-band amplitudes and phases are free parameters, the $\chi^2$ minimization procedure allows for the template fits to account for possible light-curve variations. We use the mean magnitudes corresponding to the minimum $\chi^2$ as the final mean magnitude estimates. The mean phase corrections and standard deviations for the phase corrections for {\it F555W}, {\it F814W}, and {\it F160W} are $-0.023\pm 0.210$, $-0.011\pm 0.123$, and $0.003\pm 0.070$~mag, respectively. Because the phases are free parameters, the phase errors from the PAndromeda lightcurves do not enter into the final error calculation.

\subsection{Wesenheit Indices}

We combine our {\it F555W}, {\it F814W}, and {\it F160W} mean magnitude estimates into reddening-free VIS Wesenheit ($m^W_I$) and NIR Wesenheit ($m^W_H$) indices \citep{Madore_1982} using the same formulations from \cite{Riess_2016,Riess_2019}:
\begin{eqnarray}
    m^W_I & = & m_{\textit F814W} - 1.3 (m_{\textit F555W} - m_{\textit F814W}) \\
    m^W_H & = & m_{\textit F160W} - 0.386(m_{\textit F555W} - m_{\textit F814W}) \label{eq:NIR Wesenheit}
\end{eqnarray}
\noindent where 0.386 is derived from \cite{Fitzpatrick_1999} and 1.3 is derived from \cite{Cardelli_1989ApJ...345..245C}, both using $R_V$ = 3.1. The Wesenheit indices are useful for reducing dispersion in the PLR caused by differential extinction and the non-zero temperature width of the instability strip (at a fixed period, cooler Cepheids are redder and fainter). 

For the gold sample, the final mean magnitude errors for each Cepheid are calculated by adding in quadrature the photometric errors with template errors of 0.05, 0.03 and 0.017 mag for $V$, $I$, and $H$, respectively, from \cite{Riess_2019} and \cite{Inno_2015_refId0}. An intrinsic scatter of 0.069 mag from \cite{Riess_2019} is then added in quadrature to the NIR Wesenheit index after error propagation. For the silver sample, the {\it F555W}, {\it F814W}, and {\it F160W} mean magnitude errors are calculated by adding in quadrature the photometric errors for each epoch and dividing by the square root of the number of epochs used. These errors are propagated into the VIS and NIR Wesenheit indices. We then add a random phase error of 0.13 mag from \cite{Macri_2015AJ....149..117M} in the VIS and NIR Wesenheits in quadrature. A 10$\%$ change in this random phase error only changes the final distance modulus by 2~mmag. As with the gold sample, an intrinsic scatter of 0.069 mag is additionally added in quadrature to the NIR Wesenheit index. The mean magnitude uncertainties for {\it F555W}, {\it F814W}, {\it F160W}, $m^W_I$ and $m^W_H$ after phase corrections are 0.137, 0.113, 0.073, 0.316, and 0.124 mag, respectively, for the gold sample, while for the silver sample they are 0.100, 0.076, 0.042, 0.263, and 0.164 mag, respectively.

\subsection{Geometric and Count-Rate Nonlinearity Corrections}

We applied geometric corrections to account for differences in the distance of each Cepheid from the center of M31, which can be up to a few hundredths of a magnitude. We adopt the same parameters as in \cite{Dalcanton_2012ApJS..200...18D}: an inclination of $i = 70$\textdegree, a position angle of $\theta$  = 43\textdegree, and a center of $\alpha (2000) = 10.68473$ and $\delta (2000) = 41.26805$. We estimate an error of 0.011 mag for this geometric correction by taking the standard deviation of the mean of corrections using the parameters above plus the four possible combinations of $i = 74^{\circ}$ from \cite{Courteau_2011}, $i = 77^{\circ}$ from \cite{Walterbos)1988A&A...198...61W}, $\theta$ = $45^{\circ}$ from \cite{Seigar_2008_10.1111/j.1365-2966.2008.13732.x}, and $\theta$ = $38.1^{\circ}$ from \cite{Tempel_2010}. This geometric correction is applied to $m^W_I$ and $m^W_H$, with mean values for the gold and silver samples of 0.011 mag and 0.012 mag, respectively. Mean magnitudes and their geometric corrections for each Cepheid can be found in Table \ref{tab:PhotResults}. A positive geometric correction corresponds to a Cepheid farther than the center of M31.

We also applied corrections to account for Count-Rate Nonlinearity (CRNL) or reciprocity failure, which dims fainter sources more than brighter ones due to a decreased photon collection efficiency at low count rates compared to high count rates for WFC3-IR. We adopt the CRNL correction of 0.0077 mag/dex from \cite{Riess_2019_CRNL_wfc..rept....1R}. To account for CRNL when using LMC as an anchor, as M31 is fainter than the LMC, the {\it F160W} (or equivalently, $m^W_H$) magnitudes for M31 must contain a correction corresponding to 2~dex in the brighter direction relative to the LMC. This would be equivalent to instead applying the correction so that the LMC has a 2~dex correction in the fainter direction if no correction to M31 were made. As the LMC anchor from \cite{Riess_2019} used to determine the distance to M31 below already accounts for a 4~dex CRNL correction in the fainter direction, we add $0.0154\pm0.0055$~mag to $m^W_H$ so that the LMC has a net 2~dex CRNL correction relative to M31 in the fainter direction.

\section{Period-Luminosity Relations}
\label{sec:PLR}

\begin{figure}[ht!]
\plotone{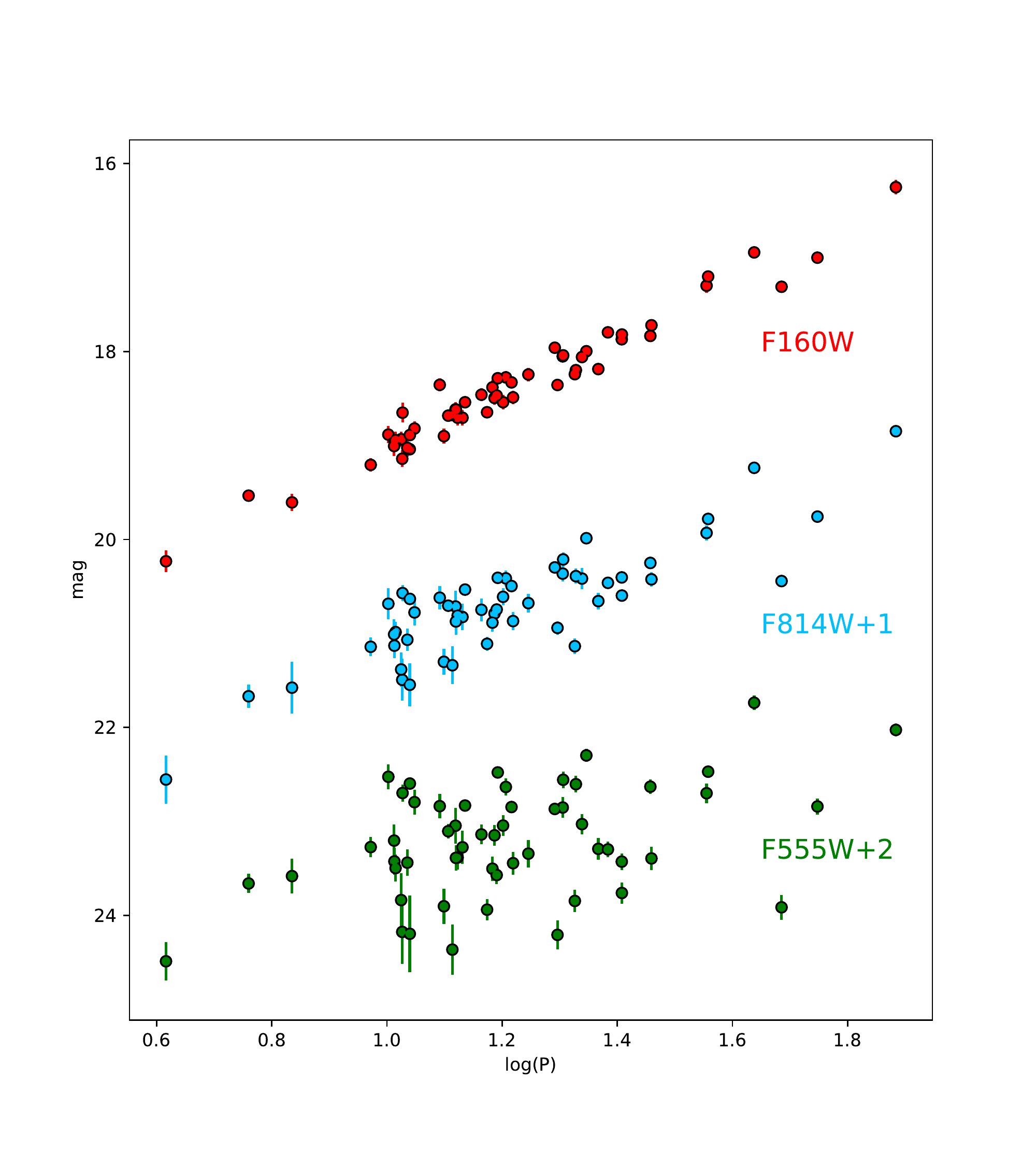}
\caption
{Period luminosity relations for {\it F555W, F814W, and F160W} in the gold+silver sample. Geometric corrections are not applied to {\it F555W, F814W, and F160W} in this plot. As explained in Section \ref{sec:PLR}, we do not attempt to fit these PLR due to high levels of differential extinction making PLR that are uncorrected for reddening highly unreliable.}
\label{fig:PLR_No_Wesen}
\end{figure}

\begin{figure}[ht!]
\plotone{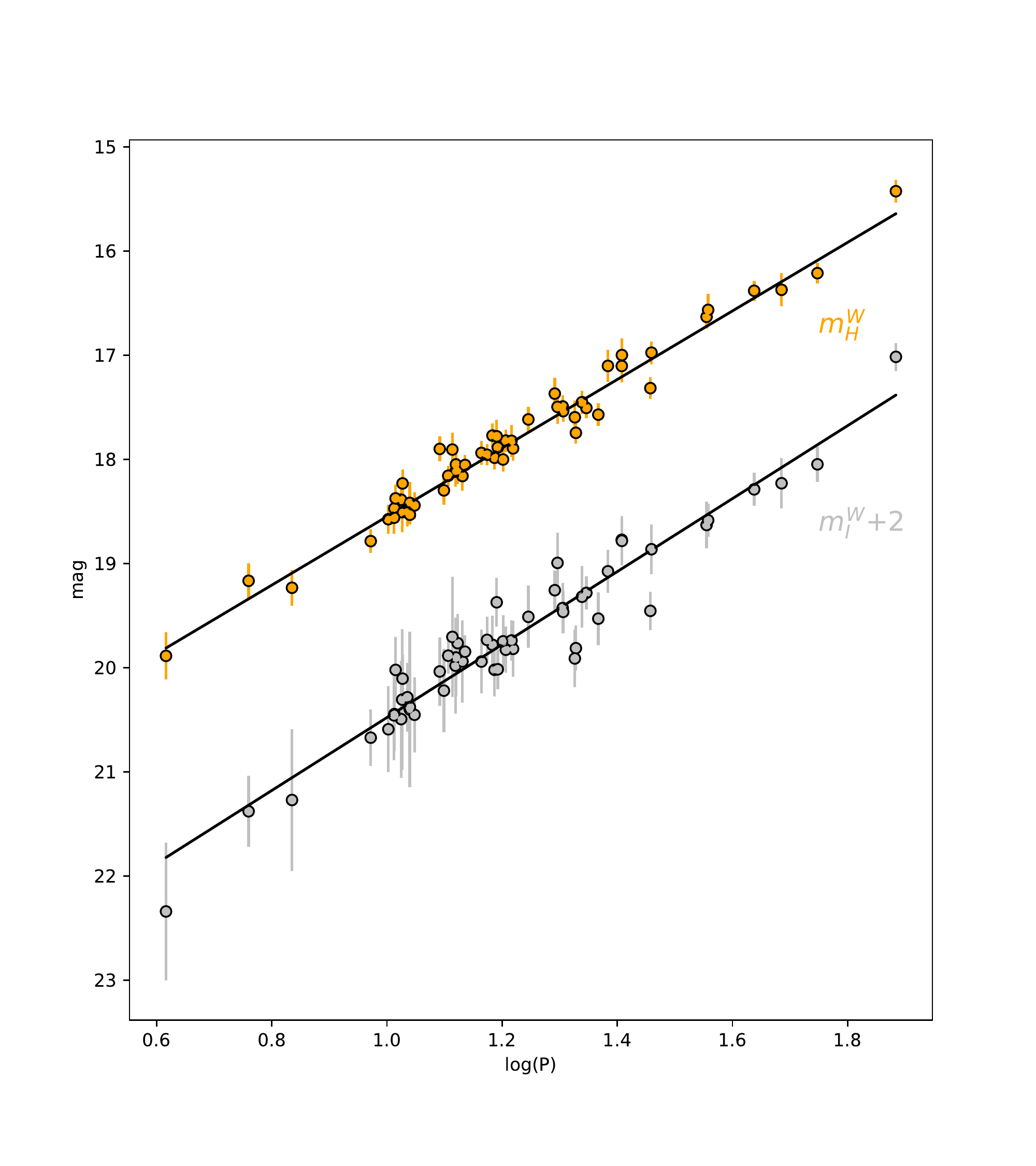}
\caption
{Period luminosity relations for near-infrared and visible Wesenheit indices in the gold+silver sample. PLR fit paramters can be found in Section \ref{sec:PLR}.}
\label{fig:PLR_Wesen}
\end{figure}

Fig. \ref{fig:PLR_No_Wesen} shows PLRs for {\it F555W}, {\it F814W}, and {\it F160W}, while Fig. \ref{fig:PLR_Wesen} presents PLRs for $m^{W}_{I}$ and $m^{W}_{H}$, both for the gold+silver sample. Weighted linear least-squares fit parameters for $m^{W}_{I}$ and $m^{W}_{H}$ for various samples can be found in Table \ref{tab:PLRFits}. Column (1) lists the Wesenheit index ($m^{W}_{I}$ or $m^{W}_{H}$), (2) corresponding sample, (3) the number of Cepheids in that sample, (4) the dispersion of a two-parameter weighted linear least-squares fit not accounting for intrinsic scatter, (5) the slope for the two-parameter fit with its error in parentheses, (6) the intercept for the two-parameter fit with its error in parentheses, (7) the chi-squared per degree of freedom, $\chi^2_{\rm dof}$, for the two-parameter fit, (8) the intercept for a weighted linear least-squares fit with fixed slopes of $-3.31$ for $m^W_I$ and $-3.26$ for $m^W_H$ \citep{Riess_2019} with its error in parenthesis, and (9) the $\chi^2_{dof}$  for the fixed-slope fit. One Cepheid (CEPH-11.12919+41.61338) in the gold sample deviates from the $m^W_H$ PLR fit beyond the value allowed by Chauvenet's criterion. We denote the sample without this object as gold+silver-1 and find small improvements in the resulting fits.

We attribute the differences in {\it F555W}, {\it F814W}, and {\it F160W} slopes compared to those in other studies \citep{Riess_2019, Wagner-Kaiser_2015_10.1093/mnras/stv880, Kodric_2018} to the effects of high differential extinction in M31. In Fig. \ref{fig:CephLocations}, we color the locations of the target Cepheids using red, white, and green for mean-magnitude residuals of $e\le -0.5, -0.5 < e < 0.5,$ and $e>= 0.5$ when compared to the {\it F555W} PLR fit from \cite{Riess_2019}, respectively. \cite{Riess_2011ApJ...730..119R} noted that young, long-period Cepheids tend to be concentrated in the spiral arms of its host galaxy where the bulk of star formation occurs while older, short-period Cepheids do not have a preferential spatial distribution. As the spiral arms contain a higher amount of dust relative to the inner portions of the disk and bulge when viewing a galaxy with high inclination such as M31 ($\sim$ 75\textdegree), long-period Cepheids in the spiral arms suffer on average from more extinction than shorter period Cepheids. As a result, the long-period side of the PLR fit is pushed faintwards more so than than the short-period side, with the relative difference dependent on the distribution of Cepheids in a given sample. The high inclination of M31 increases all extinction effects, and the increase in the difference is likely a natural consequence of this. As observations in the near-infrared are less affected by extinction, we see that there is less of a difference between the {\it F160W} slopes from this study, \cite{Wagner-Kaiser_2015_10.1093/mnras/stv880}, and \cite{Kodric_2018}, and the $m^{W}_{H}$ between this study and \cite{Riess_2019}. We find a mean reddening $R_v(E(V-I))$ for {\it F555W} of $1.577\pm 1.100$~mag using $R_V = 3.1$. We conclude that the M31 PLR dispersions and slopes by themselves are not reliable due to their dependence on period and position distribution and the effects of high differential extinction. In Fig. \ref{fig:PLR_No_Wesen}, we do not attempt to fit the PLR. However, we do fit the PLR for the Wesenheit indices in Fig. \ref{fig:PLR_Wesen} which have slopes very close to those from \cite{Riess_2019} demonstrating the effectiveness of these formulations at removing reddening even in areas of substantial differential extinction.  

\begin{deluxetable}{cccccccccc}
\tablecaption{Period-Luminosity Relation Fits to {\it HST} M31 Cepheids\label{tab:PLRFits}}
\tablehead{\multicolumn{1}{p{1.5cm}}{\centering Index} & \multicolumn{1}{p{1.5cm}}{\centering Sample} & \multicolumn{1}{p{1cm}}{\centering $N$}  & \multicolumn{1}{p{1.5cm}}{\centering Dispersion} & \multicolumn{1}{p{2cm}}{\centering Slope}& \multicolumn{1}{p{2cm}}{\centering Intercept} & \multicolumn{1}{p{2cm}}{\centering $\chi^2_{\rm dof}$} & \multicolumn{1}{p{2cm}}{\centering Intercept$^a$} & \multicolumn{1}{p{2cm}}{\centering $\chi^{2}_{\rm dof}$}}
\startdata
$m^{W}_{I}$ & Gold & 42 & 0.199 & -3.500 (0.144) & 21.990 (0.193) & 0.892 & 21.740 (0.038) & 0.913 \\
$m^{W}_{H}$ & Gold & 42 & 0.132 & -3.279 (0.082) &  21.838 (0.103) & 1.367 & 21.815 (0.018) & 1.335  \\
$m^{W}_{I}$ & Silver & 13 & 0.313 & -3.486 (0.320) & 21.929 (0.433) & 1.494 & 21.693 (0.062) & 1.395 \\
$m^{W}_{H}$ & Silver & 13 & 0.111 & -3.283 (0.181) & 21.758 (0.238) & 0.537 & 21.728 (0.045) & 0.494\\
$m^{W}_{I}$ & Gold+Silver & 55 & 0.231 & -3.501 (0.132) & 21.980 (0.177) & 0.990 & 21.727 (0.032) & 1.011\\
$m^{W}_{H}$ & Gold+Silver & 55 & 0.131 & -3.289 (0.075) & 21.839 (0.095) & 1.204 & 21.803 (0.017) & 1.184 \\
$m^{W}_{I}$ & Gold+Silver-1 & 54 & 0.232 & -3.506 (0.132) & 21.987 (0.178) & 0.987 & 21.728 (0.033) & 1.029\\
$m^{W}_{H}$ & Gold+Silver-1 & 54 & 0.124 & -3.311 (0.075) & 21.873 (0.095) & 1.061 & 21.809 (0.017) & 1.050
\enddata
\tablecomments{$m^{W}_{H}$ includes addition of $0.0154 \pm 0.0055$~mag to correct for a CRNL of 2 dex between LMC and M31 Cepheids and an intrinsic scatter term of 0.069 from \cite{Riess_2019} added in quadrature to the NIR Wesenheit mean magnitude errors. (a) Linear least squares fit using a fixed slope of -3.31 for $m^W_I$ and $-3.26$ for $m^W_H$, both from \cite{Riess_2019}.}
\end{deluxetable}

\section{Distance to M31}
\label{sec:Distance}
\subsection{NIR Distance Determination}

To derive a distance to M31, we first find the relative distance modulus between M31 and the LMC by fitting the $m^{W}_{H}$ PLR found here with a fixed slope of -3.26 from Table 3 of \cite{Riess_2019} and subtracting the resulting $m^{W}_{H}$ intercept from the $m^{W}_{H}$ intercept of 15.915, from Table 5 of \cite{Yuan_2020ApJ...902...26Y} which corrects the value from \cite{Riess_2019} for CRNL. We then add this relative distance modulus to the most direct and precise geometric distance determination to the LMC available to date of  $\mu_0=18.477\pm 0.007$~mag \citep{Pietrzynski_2019} based on detached eclipsing binaries. As \cite{Riess_2019} use the same filters as this study, we do not need to apply further corrections. Due to the wide positional spread of Cepheids over the disc of M31 for our samples, we used the deprojected metallicity gradient of $12+\log[O/H]=9.10 (\pm 0.06) -0.021 (\pm 0.0048)$~dex~kpc$^{-1}$ measured by \cite{Sanders_2012} from 60 HII regions on the same scale as \cite{Zartsky_1994ApJ...420...87Z}. These values range from $\log[O/H] = 8.72$ to 9.04, having a mean of 8.85 and a dispersion of 0.06. We adopt a mean metallicity for the LMC of $-0.3$ dex (relative to 8.9 for solar) used in \cite{Riess_2019}, as the mean of $-0.33$ dex from \cite{Romaniello_refId0} and $-0.27$ dex from \cite{Choudhury_10.1093/mnras/stv2414}. Using the metallicity term of $-0.17$ $\pm$ 0.06 mag/dex fit by \cite{Riess_2019}, we apply a $+0.04$~mag correction to the distance modulus. We thus find a distance modulus to M31 of $\mu_0= 24.419\pm 0.035$~mag ($D=765\pm 12$~kpc) for the gold sample, $\mu_0= 24.332 \pm 0.054$~mag ($D=735\pm 18$~kpc) for the silver sample, $\mu_0 = 24.407\pm 0.032$~mag ($D=761\pm 11$~kpc) for the gold+silver sample, and $\mu_0 = 24.414\pm 0.032$~mag ($D=763\pm 11$~kpc) for the gold+silver-1 sample. Although the removal of the outlier Cepheid in gold+silver-1 decreased the $\chi^2_{dof}$ to the PLR fit, it only changed the final distance by 0.007 mag, well below the final uncertainty. We find that all samples agree to $1\sigma$. The silver sample has a greater final uncertainty than the gold sample primarily due to a smaller sample size. The ratio between the PLR mean uncertainties of the silver and gold samples is given by $\sqrt{42} / \sqrt{13} = 1.80$, which is greater than the ratio between the mean $m^{W}_{H}$ magnitude uncertainties of the gold and silver samples of 1.31. For the discussion below, we refer only to the distance of gold+silver sample.

There is an ongoing discussion about the presence of a broken PLR slope at P = 10 days \citep{Sandage_2004}. Using statistical tests on LMC Cepheids, \cite{Ngeow_2008A&A...477..621N} find a broken slope in the $B, V, I_c, J$, and $H$ bands, but a linear slope in the $K_s$ and Wesenheit indices. \cite{Inno_2013} find linear NIR Wesenheit slopes for LMC Cepheids. \cite{Kodric_2018} find a broken slope for M31 Cepheids in $r, i, g$, and Wesenheit bands. However, \cite{Riess_2016} did not find a broken slope in their $m^W_H$ relation and \cite{Kodric_2018b} did not find a broken PLR slope in any of their samples using 522 fundamental mode Cepheids and 102 first overtone Cepheids in $F160W$ and $F110W$ as well as 559 fundamental mode Cepheids and 111 first overtone Cepheids in $F814W$ and $F475W$, which is the largest Cepheid sample set compiled to date. To investigate the effect of Cepheids with P $<$ 10 days on our distance determination, we removed Cepheids GRP-11.25609+41.56927 (PAndromeda ID: 1068323, Period = 6.85 days), GRP-11.03915+41.30692 (PAndromeda ID: 104540, Period = 9.362 days), GRP-11.20822+41.44917 (PAndromeda ID: 237868, Period = 4.133 days) and GRP-11.24332+41.50926 (PAndromeda ID: 1052211, Period = 5.751 days) from the gold+silver sample. We find that excluding Cepheids with P $<$ 10 days decreases the final distance modulus slightly to 24.402 $\pm$ 0.033 mag ($D = 759 \pm 11$ kpc) and does not significantly affect our final result.

A breakdown of the error budget for this distance can be found in Table \ref{tab:Error_Budget}. Column (1) lists the type of error, (2) the value of error in percent, and (3) the source where the error was derived. We combine the error in the M31 PLR mean (0.0169 mag), LMC PLR mean from \cite{Riess_2019} (0.0092 mag), count rate nonlinearity (CRNL) across 2 dex from \cite{Riess_2019_CRNL_wfc..rept....1R} (0.0013 mag), detached eclipsing binary calibration to the LMC from \cite{Pietrzynski_2019} (.0260 mag) and M31 geometry (0.0111 mag) assuming 20$\%$ uncertain. We improve upon past distance determinations to M31 by reducing the total uncertainty to 1.49$\%$, which is the lowest uncertainty published to date that includes anchor uncertainties.

\begin{deluxetable*}{ccc}
\tablecaption{Systematic Error Budget for the M31 Distance Ladder Calibration\label{tab:Error_Budget}}
\tablewidth{0pt}
\tablecolumns{3}
\tablehead{\multicolumn{1}{p{3cm}}{\centering Error\\ (1)} & \multicolumn{1}{p{2cm}}{\centering Value\\ (2)} & \multicolumn{1}{p{3cm}}{\centering Source\\ (3)}}

\startdata
 M31 P-L Mean & 0.78$\%$ & measured here\\
LMC P-L Mean & 0.42$\%$ & \cite{Riess_2019}\\
CRNL across 2 dex & 0.06$\%$ & \cite{Riess_2019_CRNL_wfc..rept....1R}\\
LMC DEB Mean & 1.20$\%$ & \cite{Pietrzynski_2019}\\ 
M31 Geometry & 0.10$\%$ & measured here \\ \hline
Total & 1.49$\%$ & 
\enddata

\end{deluxetable*}

\subsection{Comparison with Other Studies}

For a historical comparison, our determination agrees to within $1\sigma$ with the value of $\mu_0= 24.33\pm 0.12$~mag to Baade's field I (3 kpc from the center of M31) from \cite{Freedman_1990}. Our distance also agrees with all six mean estimate distances (MED) \citep{Steer_2020} and the MED for Cepheid distances only of $790\pm 78$~kpc from the NASA/IPAC Extragalactic Database (NED) (I. Steer, 2020, private communication) as well as the the recommended value of $\mu_0= 24.45\pm 0.10$~mag from \cite{de_Grijs_2014}. 

In Table \ref{tab:Recalibrated_Ceph_M31_Distances}, we compare our distance to seven highly-cited Cepheid distance measurements to M31 with less than a 10\% uncertainty. We find that our measurement agrees to within $1\sigma$ with all originally published measurements except for \cite{Saha_2006ApJS..165..108S}. However, we note that \cite{Saha_2006ApJS..165..108S} used a LMC calibration of of $\mu_0 = 18.54$~mag and that updating this calibration with the most direct and precise measurement currently available from \cite{Pietrzynski_2019} would lower their distance to $\mu_0 = 24.48$~mag and into agreement with our distance. We combine the $H,$ $I,$ and $V$ distances from \cite{Macri_2001ApJ...549..721M} into a $m^W_H$ relation using Equation \ref{eq:NIR Wesenheit}.

\begin{figure}[ht!]
\plotone{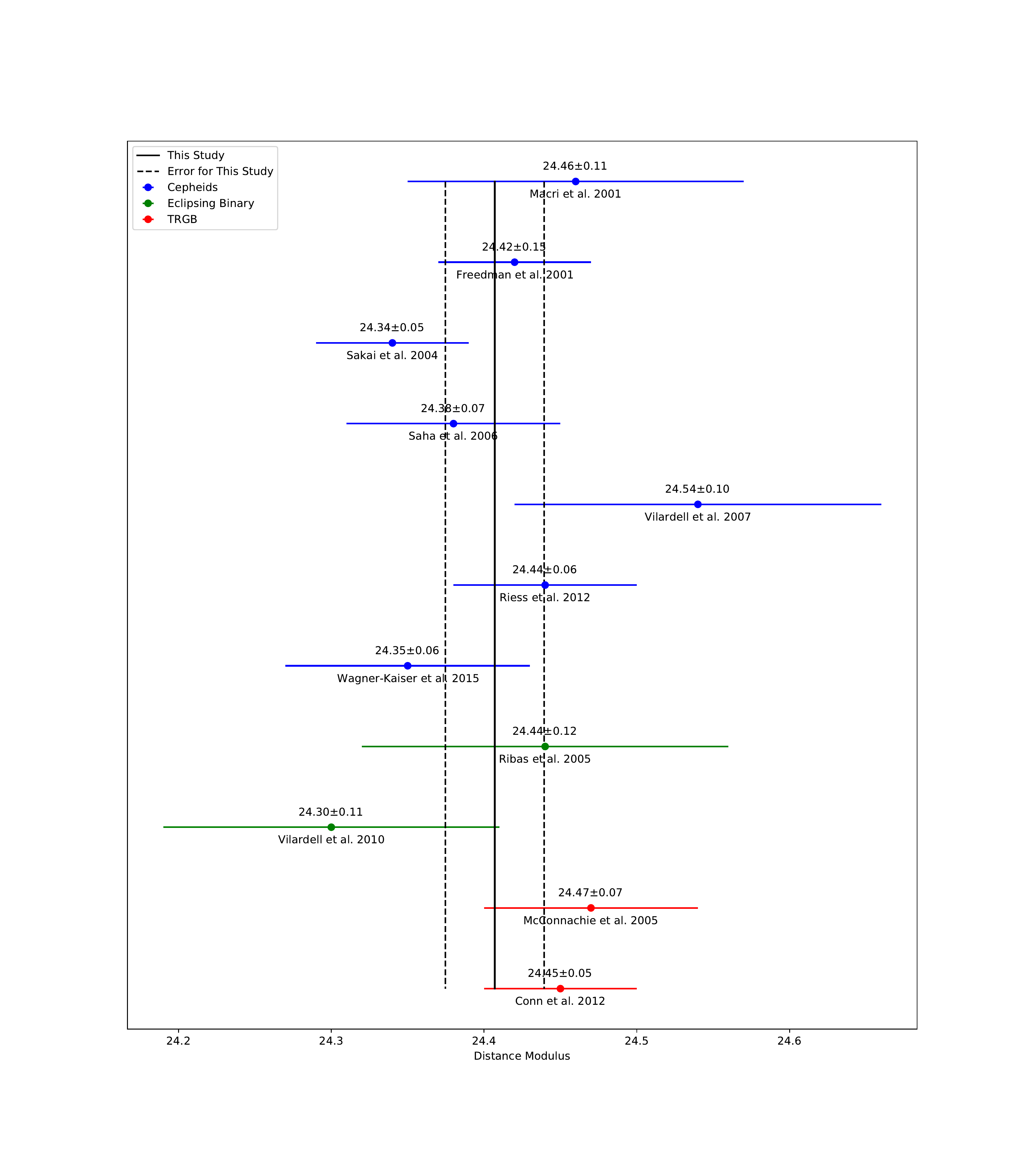}
\caption 
{Seven Cepheid, two TRGB, and all two eclipsing binary distances to M31. Cepheid measurements were recalibrated using either the DEB distance to the LMC from \cite{Pietrzynski_2019} or the water maser distance to NGC 4258 from \cite{Reid_2019}. TRGB measurements were recalibrated to a zero point of $M_{TRGB}$ = -4.05 mag. We also update metallicity corrections of -0.24 mag/dex for VIS bands from \cite{Sakai_2004ApJ...608...42S} and -0.17 mag/dex for NIR bands from \cite{Riess_2019}. We restrict Cepheid distances to those with less than 10$\%$ uncertainties. For TRGB distances, we selected those satisfying the criteria for a robust TRGB detection from \cite{Madore_1995AJ....109.1645M} and exclude distances to the outer halo or giant stellar stream. The gold+silver distance found here is shown by the vertical black line with 1$\sigma$ errors represented by the dashed vertical black lines.}
\label{fig:M31DistancesRecalibrated}
\end{figure}

As calibrations typically improve over time, in Table \ref{tab:Recalibrated_Ceph_M31_Distances} and Fig.~\ref{fig:M31DistancesRecalibrated} it is reasonable to recalibrate the seven prior M31 Cepheid distance measurements with more recent measurements of their geometric distance calibration or Cepheid metallicity dependence to provide a more homogeneous comparison. In Table \ref{tab:Recalibrated_Ceph_M31_Distances}, column (1) lists the Cepheid study, (2) the original published distance in the top row and the distance after recalibration in the bottom row, (3) the original anchor galaxy in the top row and the recalibration distance reference (given in the table notes) in the bottom row,  (4) the original anchor distance and uncertainty, (5) the original metallicity correction in the top row and the updated metallicity correction in the bottom row, and (6) the original metallicity correction in mag/dex. We recalibrate the metallicity corrections to match the most recent corrections of $-0.17$~mag/dex from \cite{Riess_2019}  for NIR bands and $-0.24$~mag/dex from \cite{Sakai_2004ApJ...608...42S} for VIS bands, and a metallicity difference of $\log[O/H] = 8.85 - 8.60 = 0.25$~between the LMC and M31. For comparison, Fig. \ref{fig:M31DistancesRecalibrated} also includes two TRGB and all two eclipsing binary measurements, all of which agree to within 1$\sigma$. We also calibrate the two TRGB distances to a common zeropoint of $M_{\textrm TRGB}= -4.05$~mag used in \cite{Freedman_2019} to compare Cepheid and TRGB distances, discussed further in Section \ref{sec:CepheidvsTRGB}.

\begin{deluxetable*}{cccccc}
\label{tab:Recalibrated_Ceph_M31_Distances}
\tablecaption{Recalibrated Cepheid Distances\label{tab:Fits}}
\tablehead{\multicolumn{1}{p{3.5cm}}{\centering Study \\}  & \multicolumn{1}{p{3cm}}{\centering Published $\mu_0$ \\ (Updated)} & \multicolumn{1}{p{1cm}}{\centering Anchor} & \multicolumn{1}{p{2.5cm}}{\centering Anchor $\mu_0$ (mag)\\ (Updated) } & \multicolumn{1}{p{2.5cm}}{\centering Metal. Corr. (mag)\\ (Updated)$^g$} & \multicolumn{1}{p{2cm}}{\centering Metal. Corr.\\ (mag/dex)} 
}

\startdata
\cite{Macri_2001ApJ...549..721M} &  \ \ \ \ 24.44 (0.11)$^{a,b}$ & LMC & 18.50 (0.10) & 0 & 0 \\
 &  \ \  \ \  24.46 (0.11) & & (1)  & 0.04 &  \\ \\
\cite{Freedman_2001ApJ...553...47F} &  \ \ \ \ 24.48 (0.05)$^a$ & LMC & 18.50 (0.10) & 0.10 & -0.2 \\
&  \ \ \ \ 24.42 (0.05) & & (1)  &  0.06&    \\ \\
\cite{Sakai_2004ApJ...608...42S} &   \ \ \ \ 24.38 (0.05)$^a$ & LMC & 18.50 (0.10) & 0.08$^e$ & -0.24 (0.05)  \\
&    \ \ \ \  24.34 (0.05)& & (1)  & 0.06 & \\ \\
\cite{Saha_2006ApJS..165..108S} &  \ \ \ \  24.54 (0.07)$^a$ & LMC & 18.54$^c$ & 0.16 & -0.39 (0.03)\\
&    \ \ \ \  24.38 (0.07)  & &(1)  & 0.06 &  \\\\
\cite{Vilardell_2007refId0} &   \ \ \ \  24.32 (0.12) & LMC & 18.42 (0.06) & -0.10$^f$ & -0.25 (0.09) \\
&    \ \ \ \  24.54 (0.10) & & (1) &  0.06 & \\\\
\cite{Riess_2011} &  \ \ \ \ 24.38 (0.064) & MW & -5.86 (0.04)$^d$ & 0 & 0 \\
&    \ \ \ \ 24.44 (0.06) & & (3) &  0&  \\ \\
\cite{Wagner-Kaiser_2015_10.1093/mnras/stv880} &   \ \ \ \ 24.51 (0.08) & NGC 4258 & 29.40 (0.06) & 0.16 & 0.22 (0.60) \\ 
&    \ \ \ \ 24.35 (0.06) & & (2) & 0 &   \\ &  & &  &&\\ \hline
This Study &   \ \ \ \  24.41 (0.03)  & LMC & (1) & & 
\enddata

\tablecomments{(1) $\mu_0(LMC) = 18.477 \pm 0.007$~mag \citep{Pietrzynski_2019}, (2) $\mu_0(N4258)= 29.397 \pm 0.031$~mag \citep{Reid_2019}, (3) $M^W_H = -5.91 \pm 0.022$~mag \citep{Riess_2021}. (a) Anchor error was not included in the original distance measurement. (b) $H$-, $I$-, and $V$-band distances combined into NIR Wesenheit distance using Eq.~\ref{eq:NIR Wesenheit}. (c) Error not reported in original paper. (d) Absolute magnitude (e) Original metallicity correction was calculated using $\Delta\log[O/H] = 8.68 - 8.34 = 0.34$ from Table 4, Column 3. (f) Metallicity corrections were applied individually to each Cepheid. A table of these corrections was not available, so we estimate the mean correction as the average of -0.15 mag and -0.05 mag as stated in \cite{Vilardell_2007refId0}, \S 5.2. (g)  Updated metallicity corrections adopt the most recent corrections of $-0.17$~mag/dex from \cite{Riess_2019} for NIR bands and $-0.24$~mag/dex from \cite{Sakai_2004ApJ...608...42S} for VIS bands, and a metallicity difference of $\log[O/H] = 8.85 - 8.60 = 0.25$ between the LMC and M31.} 
\end{deluxetable*}

\subsection{Cepheid and TRGB Comparison}
\label{sec:CepheidvsTRGB}
There is a well-known discrepancy between indirect and direct determinations of the Hubble constant, dubbed the ``Hubble Tension". However, there is also a smaller difference between the Cepheid and TRGB determinations which differ at the 4\% level \citep{Riess_2021, Freedman_2019} with $<2\sigma$ significance. The availability of both measures at high precision for M31 allows us to further investigate this difference.

The nearness of M31 and sparseness of TRGB stars necessitates a wide-angle survey of the halo of M31 to produce a precise measure of its TRGB.  Unfortunately, the few M31 halo fields observed with {\it HST} do not produce a well-sampled, reliable determination of the TRGB.  To ensure a meaningful comparison we therefore select ground-based studies that measure the TRGB in the $I$ band or equivalent and satisfy the criterion for a meaningful detection of the TRGB from \cite{Madore_1995AJ....109.1645M}, which requires a minimum number of 100 stars in the bin one magnitude fainter than the identified TRGB. There are two such ground-based studies that satisfy these criteria.   \cite{McConnachie_2005} used a 40 deg$^2$ survey with the Isaac Newton Telescope and assuming $M_I= -4.05$ found $\mu_0=24.47 \pm  0.07$~mag. In addition, \cite{Conn_2012ApJ...758...11C}  used the PAndAS survey undertaken on the Canada-France-Hawaii Telescope and, assuming $M_I=-4.04$ (transformed in \cite{Bellazini_2008MmSAI..79..440B}), derived $\mu_0=24.46\pm 0.05$~mag. These results are in good agreement with earlier studies from \cite{Sakai_2004ApJ...608...42S} and \cite{Ferrarese_2000ApJ...529..745F} both of which result in a lower precision due to sparse color magnitude diagrams from \cite{Mould_1986ApJ...305..591M}. 

In Table \ref{tab:TRGB}, we provide the results from \cite{McConnachie_2005} and \cite{Conn_2012ApJ...758...11C} used to measure distances to M31 from their TRGB determinations. Column (1) lists the TRGB study, (2) the original derived distance modulus $\mu_0$ (3) the original value of $M_I$, (4) the assumed extinction, and (5) the implied change in $M_I$ by using the distance derived in this study to recalibrate the TRGB luminosity.

Nominally, the distance to M31 determined here from Cepheids on the HST system in M31 and the LMC and the LMC DEB distance is $\sim$ 0.05 mag closer than the best estimates derived from TRGB with the calibration of $M_I=-4.05$ mag, implying $M_I=-4.00$ mag if we attribute the difference to the TRGB calibration and not chance\footnote{We would not know how to apply this difference directly to the Cepheid luminosity because we derive the distance to M31 from a direct comparison to LMC Cepheids rather than from an independent determination of Cepheid luminosity.}  We take a wider view of this difference in the Discussion by including the consideration of other hosts.

\begin{deluxetable*}{ccccc}
\label{tab:Recalibrated_M31_Distances}
\tablecaption{Recalibrated TRGB Zeropoints\label{tab:TRGB}}
\tabletypesize{\footnotesize}
\tablecolumns{5}
\tablehead{\multicolumn{1}{p{4cm}}{\centering TRGB Study \\ (1)} & \multicolumn{1}{p{3cm}}{\centering Original $\mu_0$\\ (2)} & \multicolumn{1}{p{2.5cm}}{\centering Original $M_I$\\ (3)}& \multicolumn{1}{p{1cm}}{\centering$A_I$\\ (4)}& \multicolumn{1}{p{2cm}}{\centering $\Delta M_I$\\ (5)}}
\startdata
\cite{McConnachie_2005} & 24.47 (0.07) & -4.05 (0.05) & 0.12$^a$ & 0.06 \\
\cite{Conn_2012ApJ...758...11C} & 24.46 (0.05) & -4.04 (0.12) & 0.12$^a$& 0.05
\enddata
\tablecomments{ \textsuperscript{a}Converted from E(B-V) using $R_I = A_I/E(B-V) = 1.94$ from \cite{Schlegel_1998ApJ...500..525S}}
\end{deluxetable*}

\subsection{M31 as a Potential Anchor}

While there are currently two geometric non-detached eclipsing distances to M31 that could be used to turn M31 into an anchor galaxy, both are from early-type stars \citep[for instance, late O and early B spectral types from][] {Ribas_2005} and rely on non-local thermal equilibrium models that are not strictly geometric and do not yet have a robust grasp on uncertainties. In addition, non-detached eclipsing measurements can be affected by distortion and reflection \citep{Vilardell_2010A&A...509A..70V} that rely on non-empirical models to correct, and the use of these eclipsing binaries may differ from cooler, late-type stars \citep{de_Grijs_2014}. While precise empirical calibrations exist for late-type DEBs \citep{Pietrzynski_2019} and currently hundreds of eclipsing binaries have been discovered in M31 \citep[e.g.,][]{Lee_2014}, only 11 are bright enough ($V<20.5$~mag) for radial-velocity observations and would require observations with large, 8-10 meter ground-based telescopes to obtain radial velocity measurements with suitable signal-to-noise ratios \citep{Beaton_2018SSRv..214..113B}. This may be feasible with forthcoming larger telescopes.

In addition, while five water masers have been discovered in M31 by \cite{Darling_2011ApJ...732L...2D}, only two are bright enough for observations with the Very Long Baseline Interferometer and both reside in high mass star forming regions that have high noncircular motions which restrict precision of a geometric distance to $\sim$ 20$\%$ (A. Brunthaler, 2021, private communication). A late-type detached eclipsing binary or additional bright water maser measurement in M31 could help tie the Cepheid PLR here to a robust anchor that would allow M31 to be used as a high precision anchor on par with the MW, LMC, and NGC 4258 and further reduce uncertainty in the Hubble constant by diversifying anchors. 

\section{Discussion}
\label{sec:Discussion}

We construct Cepheid PLRs in M31 in {\it F555W}, {\it F814W}, and {\it F160W}, as well as VIS and NIR Wesenheit indices constructed from these filters. We also improve the uncertainty in the Cepheid distance to M31 to the 1.5$\%$ percent level, providing high precision groundwork for using M31 as an anchor galaxy in the cosmic distance ladder that is comparable to other anchor galaxies such as the LMC and NGC 4258. Despite the unfortunate degradation of \textit{HST} Gyro 2 during the GO-15146 program and the loss of several images, we were still able to significantly reduce the uncertainties from past studies by using the same filter system and instrument as our LMC calibration as well as using a sample set located at the front of M31 that reduces crowding.

While not statistically significant, the $\sim0.05$~mag difference between TRGB and Cepheids measured in the previous section is in the same direction and similar in size as the 0.059 mag weighted mean offset identified by \cite{Freedman_2019} from their comparison of 28 galaxies with Cepheid and TRGB distances, for which the TRGB distances were also calibrated to $M_I=-4.05$~mag.  However, \cite{Freedman_2019} do not provide the identity of these galaxies, the significance of this difference or the source of the distance measurements, so we cannot draw strong conclusions from this comparison. 

The {\it HST} Key Project presented a calibration of TRGB by directly comparing 9 hosts of Cepheid and TRGB measurements in \cite{Ferrarese_2000ApJ...529..745F} and found $M_I=-3.99 \pm 0.07$~mag assuming an optical Cepheid metallicity correction term of $-0.24$~mag/dex \citep{Sakai_2004ApJ...608...42S}, and $-4.06 \pm 0.07$~mag in the absence of a metallicity correction. The same metallicity term was used in the determinations of $H_0$ by \cite{Freedman_2001ApJ...553...47F,freedman_2012ApJ...758...24F}. This correction has been recently confirmed to higher precision by \cite{Breuval_2021arXiv210310894B} who found $-0.251\pm 0.057$~mag/dex between the MW, LMC and the Small Magellanic Cloud (SMC). \cite{Kourkchi_2020} also find a discrepancy between Cepheid and TRGB distance measurements in the same direction, a 0.13 mag disagreement for 25 galaxies. Thus at face value, a concordance between Cepheid and TRGB distances appears to occur at a TRGB calibration of $M_I \sim -4.00$ which is 0.05 mag fainter than the \cite{Freedman_2019} calibration. This difference does not appear to depend on distance and the distance range of the comparison $D<20$ Mpc is within the range where both methods are well measured. 

A fainter calibration of TRGB, $M_I=-3.97\pm 0.06$~mag, was found by \cite{soltis2020parallax} from the Gaia EDR3 parallax of $\omega$~Cen \citep[which is consistent with][]{Baumgardt2019MeanPM, Capozzi_PhysRevD.102.083007,Maiz_2021arXiv210110206M}.  The size and direction of this difference in calibration is also seen in two recent calibrations of TRGB in the same galaxy,  NGC 4258, using the same maser distance by \cite{Reid_2019}.   \cite{Jang_2021} find $M_I=-4.05$ and \cite{Anand_2021_10.1093/mnras/staa3668} report $M_I=-3.99$.  These studies use different methods to measure the apparent location of the TRGB in NGC 4258.  \cite{Jang_2021} find an apparent magnitude of $m_{\textit F814W} = 25.38\pm 0.02$ using edge detection with a Sobel filter for the same image (GO 9477, ACS) for which the Extragalactic Distance Database approach from \cite{Anand_2021_10.1093/mnras/staa3668} using maximum likelihood (fitting a broken luminosity function) yields $25.43\pm 0.03$~mag. Differences in the way stars are selected may play a role. The LMC is a more challenging place to refine the calibration of the TRGB due to the presence of significant foreground extinction, which is not present to the same level in the halos of other galaxies where the TRGB is measured. This makes it necessary to estimate and remove the absolute extinction rather than relative extinction which suffers less bias due to cancellation \citep{Yuan_2019ApJ...886...61Y, Freedman_2020ApJ...891...57F, Nataf_2021ApJ...910..121N}.  The SMC is also a less ideal laboratory for calibration work due to its large depth.  Unfortunately, many of these TRGB and Cepheid comparisons do not use homogeneous data or are not well measured and no comparison may be individually highly significant making it hard to resolve this $\sim0.05$~mag difference at present. For this reason we have not revisited these comparisons as their conclusions may not be applicable to present measures. However, this difference in TRGB calibrations, most notably the 0.050 $\pm$ 0.036 mag difference in different methods to measure the brightness of the NGC 4258 tip from the same data which is the simplest to consider because it is independent of all external factors,  is of the right size and direction to explain most of the difference in each measurement of $H_0$ which is $5\log(73/70) \sim 0.09$~mag and bares attention.

A clearer resolution may come from new and strong geometric calibrations derived directly from {\it Gaia} EDR3 parallaxes.   Direct parallaxes of 75 Cepheids from {\it Gaia} EDR3 with {\it HST} photometry constrain the Cepheid calibration (of SH0ES or similarly of the Key Project, neither changed) to 1\% in distances \citep{Riess_2021}.  This calibration from {\it Gaia} EDR3 Cepheid parallaxes is supported by prior, independent parallax measurements including those measured with spatial scanning of WFC3 on {\it HST} \citep{Riess_2018}, with the FGS on {\it HST} \citep{Benedict_2007AJ....133.1810B} or by using Cepheid companions, binary or cluster \citep{Breuval_2020_refId0} in {\it Gaia} DR2. An enhanced network of common calibration enabled by {\it Gaia} and a growing set of hosts with distance measures with both methods is likely to reduce any remaining discrepancy.  In addition, new measurements of these techniques and others with the {\it James Webb Space Telescope} may provide improved calibrations.

\acknowledgments

This research has made use of the NASA/IPAC Extragalactic Database (NED),
which is operated by the Jet Propulsion Laboratory, California Institute of Technology,
under contract with the National Aeronautics and Space Administration.

MPB is supported by a National Science Foundation Graduate Research Fellowship under grant No. 1746891. MPB thanks the LSSTc Data Science Fellowship Program, which is funded by LSSTc, NSF Cybertraining Grant No. 1829740, the Brinson Foundation, and the Moore Foundation; their participation in the program has benefited this work. MPB has been supported by the Director's Research Fund at STScI.

Support for this work was provided by the National Aeronautics and Space Administration (NASA) through program No. GO-15146 from the Space Telescope Science Institute (STScI), which is operated by AURA, Inc., under NASA contract No. NAS 5-26555. A.G.R., S.C., and L.M.M. gratefully acknowledge support by the Munich Institute for Astro- and Particle Physics (MIAPP) of the DFG cluster of excellence "Origin and Structure of the Universe."

This research is based primarily on observations with the NASA/ESA Hubble Space Telescope, obtained at STScI, which is operated by AURA, Inc., under NASA contract No. NAS 5-26555.

The HST data used in this paper are available at https://archive.stsci.edu/doi/resolve/resolve.html?doi=10.17909/t9-epb2-fp05 as part of the MAST archive which can be accessed at http://archive.stsci.edu/hst/.

%




\clearpage

\clearpage

\clearpage

\clearpage




\bibliography{sample63}{}
\bibliographystyle{aasjournal}



\end{document}